\documentclass[12pt]{article}
\usepackage{graphicx}
\usepackage{amsmath}
\usepackage{amssymb}

\def\gtap{\raisebox{-.4ex}{\rlap{$\sim$}} \raisebox{.4ex}{$>$}}

\def\zp {$Z^{\prime}$}
\def\zpsp {$Z^{\prime}\ $}

\def\zpysp {$Z_Y^{\prime}\ $}

\def\sth34{$|{\rm sin}\theta_{34}|$}
\def\sth34sp{$|{\rm sin}\theta_{34}|\ $}

\def\zzpsp {$Z-Z^{\prime}\ $}

\def\alrsp {$A_{LR}\ $}

\def\afbbsp {$A_{FB}^b\ $}

\def\afbc {$A_{FB}^c$}
\def\afbcsp {$A_{FB}^c\ $}

\def\mh {$m_H$}
\def\mhsp {$m_H$\ }
\def\chisqsp {$\chi^2$\ }

\def\th34{$\theta_{34}$}
\def\th34sp{$\theta_{34}\ $}

\def\dalfive{$\Delta \alpha^{(5)}$}

\def\als{$\alpha_S$}%

\def\epemsp{$e^+e^-\ $}
\def\journal{\topmargin 0.0in   \oddsidemargin 0in
        \headheight 0pt \headsep 0pt
        \textwidth 6.5in 
\textheight 9in 
        \marginparwidth 1.5in
        \parindent 2em
        \parskip .5ex plus .1ex         \jot = 1.5ex}
\journal

\begin{document}
\begin{titlepage}

\noindent  May 4, 2011\\

\begin{center}

\vskip .5in
{\large A heavy little Higgs and a light \zpsp under the radar}
\vskip .5in

Michael S. Chanowitz

\vskip .2in

{\em Theoretical Physics Group\\
     Lawrence Berkeley National Laboratory\\
     University of California\\
     Berkeley, California 94720}
\end{center}

\vskip .25in

\begin{abstract}

The original littlest Higgs model with universal fermion couplings is
found to be consistent with precision electroweak data but is strongly
constrained by Tevatron limits on the predicted centi-weak \zpsp
boson. A possible signal observed by CDF at 240 GeV is consistent with
the predicted \zp, and a region below 150 GeV is largely unconstrained
by collider data. LHC searches for narrow dilepton resonances below
500 GeV will have sufficient sensitivity to discover the \zpsp boson
or to exclude the model over most of the range allowed by the
electroweak fits.

\end{abstract}

\end{titlepage}

\newpage

\renewcommand{\thepage}{\arabic{page}}
\setcounter{page}{1}

\noindent{\bf Introduction} 

Little Higgs models address the fine tuning problem posed by
quadratically divergent one loop corrections to the Higgs boson mass
in the SM (Standard Model) by identifying the Higgs as a
pseudo-Nambu-Goldstone boson which only acquires a cutoff sensitive
mass at two loop order. Pioneering studies of the original
$SU(5)/SO(5)$ littlest Higgs model\cite{ackn} found that constraints
from precision EW (electroweak) data\cite{chkmt1,chkmt2,cd,mss} and
collider limits on the predicted \zpsp boson\cite{hpr} force the model
into a region where fine-tuning re-emerges, engendering many variants
of the original model.\cite{LHreview} Here we present EW fits of the
original model that are consistent with the precision data and in
which the Higgs mass is not fine-tuned. Good fits, with \chisqsp
values below the SM fit and fine tuning above 10\% (and often of order
one), occur for values of the $SU(5)$ breaking condensate $f$ between
0.5 and 2.7 TeV. The best of these fits are at $f= O(1)$ TeV, as
orginally envisioned in \cite {ackn}, while unexpectedly favoring
large values of the Higgs boson mass, from $\sim 0.3 - 1$ TeV. The
model then also removes the tension between the EW data and the LEPII
lower limit on the Higgs mass, which is especially acute if the
\afbbsp anomaly is due to underestimated systematic
error.\cite{mc-lose^2}

A signature prediction of the fits is a light \zpsp boson below
$\simeq 500$ GeV and possibly as light as ${\rm O}(100)$ GeV, with
centi-weak coupling to quarks and leptons. CDF\cite{cdf} and
D0\cite{d0} limits currently provide the strongest constraints,
excluding much of the region allowed by the EW fits. An excess at 240
GeV in the $e^+e^-$ mass spectrum observed by CDF\cite{cdf} is
consistent with the \zpsp predicted by the EW fits. The excess is
nominally $3.8\sigma$, with 0.6\% probability ($2.5\sigma$) to be due
to a chance fluctuation anywhere in the 150-1000 GeV mass range.  If
confirmed as a \zpsp boson, it would correspond in the LH (littlest
Higgs) model to a symmetry breaking scale $f \simeq 1.5$ TeV, and
would provide a good fit to the EW data. The CDF and D0 studies
have comparable sensitivity, since D0 considered a larger data sample
while CDF had a larger acceptance, and the CDF excess is outside the
D0 allowed region. Future Tevatron and LHC data will soon determine if
the excess is a fluctuation or a real signal.  The model can be tuned
to further suppress \zpsp production, but without a physical basis
from the UV completion it would be strongly disfavored unless a signal
emerges at or near the present limits.

Following \cite{chkmt2} we assume universal fermion charge assignments
for the two $U(1)$ gauge groups embedded in the global $SO(5)$: the
first two SM families have the same $U(1)_i$ charges as the third
family, fixed by gauge invariance of the top quark Yukawa interaction
specifed in the original model\cite{ackn} and the absence of mixed
$SU(2)_L - U(1)_i$ anomalies.  The results differ from earlier
studies\cite{chkmt2,mss} chiefly because the EW fits are performed
with complete scans of both the SM and LH parameters, possible 
thanks to currently available computing capability. Earlier studies
fixed the SM parameters at their SM best fit values and/or did not
scan on all LH parameters.  We find that the LH best fit typically
occurs at different values of the SM parameters than the SM best fit
(especially \mh) and that important cancellations emerge if all LH
parameters are scanned. Current data is more restrictive than the data
used in earlier studies --- in addition to more precise measurements
of the top and $W$ masses, low energy data\cite{apv,moller} and
Tevatron limits on \zpsp production now impose stronger constraints.
ZFITTER\cite{zfitter} is used for the SM corrections, and experimental
correlations are included.\cite{ewwg}

In the next section we discuss the methodology of the EW fits and
summarize the results. We then discuss the light centi-weak \zpsp
boson that is predicted by the fits, including the upper bounds from
the Tevatron and the fits that result if the excess at 240 GeV seen by
CDF is attributed to the \zp. This is followed by a discussion of the
future limits that can be reached at the LHC. We next discuss the
extent to which the parameters of the model are themselves fine tuned,
concluding that the principal source of tuning is the constraint imposed
on the $U(1)$ mixing angle by the Tevatron (and eventually LHC) bounds
on \zpsp production. We conclude with a brief discussion of the
results.

\vskip 0.1 in

\noindent{\bf Electroweak Fits} 

The global $SU(5)$ contains a gauged $SU(2)_1 \times SU(2)_2 \times
U(1)_1 \times U(1)_2$ subgroup with coupling constants
$g_1,g_2,g^{\prime}_1,g^{\prime}_2$. The breaking to $SO(5)$ with
condensate $f$ gives masses to a combination of the $SU(2)_i$ and
$U(1)_i$ gauge bosons.  The orthogonal $SU(2) \times U(1)$ is unbroken
and the would-be Higgs boson is at this stage part of a massless
Nambu-Goldstone boson doublet. The unbroken $SU(2) \times U(1)$ is
identified with the EW $SU(2) \times U(1)$ and is subsequently broken
by a Higgs boson vev (vacuum expectation value), $v=247$ GeV, induced
by the one loop effective potential --- for details see \cite{ackn}
and \cite{chkmt1}.  

The salient features for the EW fit are (1) changes in $Z$ boson
interactions from \zzpsp mixing and (2) custodial $SU(2)$ breaking
from three sources: a triplet Higgs boson vev, the shift in $m_Z$
due to \zzpsp mixing, and mixing between the left chirality $t_L$
quark and its $t^{\prime}_L$ partner.  We scan the usual SM
parameters, \dalfive, \als, $m_t$, and \mh, and the LH parameters
which affect the fit: the $SU(5)$ breaking scale $f$, the triplet
Higgs vev $v^{\prime}$, the sine of the $t_L-t^{\prime}_L$ mixing
angle $s_L$, and the cosines of the $SU(2)_i$ and $U(1)_i$ mixing
angles $c$ and $c^{\prime}$, related to the SM EW couplings by
$g=sg_1=cg_2$ and
$g^{\prime}=s^{\prime}g^{\prime}_1=c^{\prime}g^{\prime}_2$. 

The universal fermion $U(1)$ charge assignments are parameterized as
$y_1 = (1-\eta^{\prime})y_{SM}$ and $y_2 = \eta^{\prime}
y_{SM}$. Gauge invariance of the Yukawa interaction proposed in
\cite{ackn} then requires\cite{chkmt2} that $\eta^{\prime} = 2/5$, and
the correction to the $Z$ coupling for fermion $f$ with SM coupling
$g_f=t_{3f} - s_W^2q_f$ is
$$
\delta g_f= \frac{v^2}{2f^2}\left\{t_{3f}\left[c^2(1-2c^2) +
         5(c^{\prime\, 2}- \eta^{\prime})(1-2c^{\prime\, 2})\right]
         -5q_f(c^{\prime\, 2}-\eta^{\prime})(1-2c^{\prime\, 2})\right\} 
           \eqno{(1)}
$$ 
where $t_{3f}$ and $q_f$ are the weak isospin and charge of fermion $f$,
$s_W^2 = {\rm sin}^2\theta_W$, and $\eta^{\prime}= 2/5$ follows from 
the universal charge assignment. Corrections to 
the low energy parameters are 
$$
s_*^2(0)= s_W^2\left\{1-{v^2 \over 2f^2}\left[ c^2 
         +5(c^{\prime\, 2}- \eta^{\prime})(1-2\eta^{\prime})
         \left(1 - {1\over s_W^2}\right)\right]\right\}
                                        \eqno{(2)}
$$
and 
$$
\delta\rho_*= {5\over 4}{v^2 \over f^2}(1-2\eta^{\prime})^2
                -4{v^{\prime\, 2}\over v^2}  \eqno{(3)}
$$ 
These results are consistent with \cite{chkmt1,chkmt2}.\footnote{Sign
  errors in eq. (3.10) of \cite{chkmt1} do not propagate to the
  the appendix of \cite{chkmt1} which we have verified.}

The fits are performed subject to three conditions. First, requiring
$|v^{\prime}| < |v^2/4f|$ ensures positivity of the triplet Higgs
mass. Second, since the coefficient $a$ of the quadratically divergent
term in the one loop gauge boson effective potential is expected to be
of order one, we require $ 1/5 <a< 5$, where $a$ is determined 
by\footnote{The potential eq. (4.16) of \cite{chkmt1}
  reverses $g_1 \leftrightarrow g_2$ and $g_1^{\prime} \leftrightarrow
  g_2^{\prime}$ relative to eq.(4.7) of \cite{ackn}; we follow
  \cite{ackn}.}
$$
a= \frac{m_H^2}{4m_Z^2}\, \frac{c^2 c^{\prime\, 2}}
                    {s_W^2 c^2 +c_W^2 c^{\prime\, 2}}\, 
          \frac{1}{1+ |4 v^{\prime}f/v^2|}.     \eqno{(4)}
$$ 
Third, following the earlier studies\cite{ackn,chkmt1,chkmt2,mss} we
consider the residual fine tuning from the top partner that cuts off
the quadratically divergent top quark contribution to \mhsp and is the
most important source of the SM little hierarchy problem.\footnote{For
  additional discussions of fine tuning in the LH model see
  \cite{ceh,gku}.} We require this residual fine tuning to be no
less than 10\%,
$$
\delta_{FT}= \frac{m_H^2}{(3m_t^2 m_t^{\prime\, 2}/2\pi^2v^2)
           {\rm ln}(4\pi f/m_{t^{\prime}})}
           > 0.1,  \eqno{(5)}
$$
where $m_{t^{\prime}}^2= m_t^2\, f/(s_Lv-s_L^2f)$. 
Following \cite{chkmt1,chkmt2} we also restrict $\theta$ and
$\theta^{\prime}$ to $s,c,s^{\prime},c^{\prime}\, \geq\, 0.1$ to keep
the gauge coupling constants from becoming unreasonably large.

\begin{figure}[t]
\centerline{\includegraphics[width=5in,angle=0]{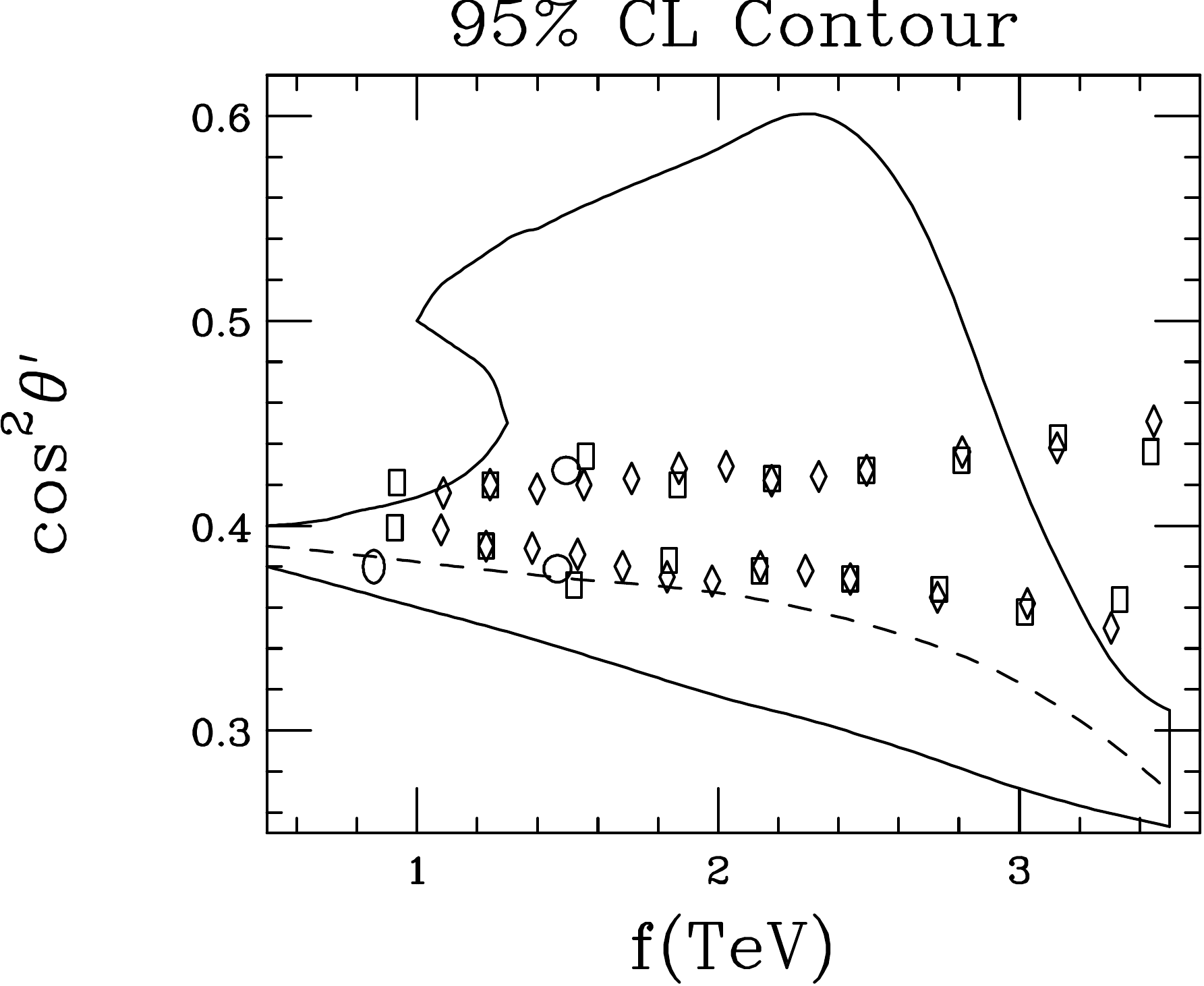}}
\caption{95\% CL contour for EW fits satisfying boundary
  conditions. The dashed line marks the best fit. Diamonds and boxes
  are upper and lower limits on $c^{\prime\, 2}$ obtained from the D0
  and CDF limits on \zpsp production, and the two circles correspond
  to the CDF excess at 240 GeV. The ellipse corresponds to a \zpsp
  boson at 140 GeV that would be unobservable at LEPII, as discussed
  in the text.}
\label{fig1}
\end{figure}

\begin{figure}[t]
\centerline{\includegraphics[width=5in,angle=0]{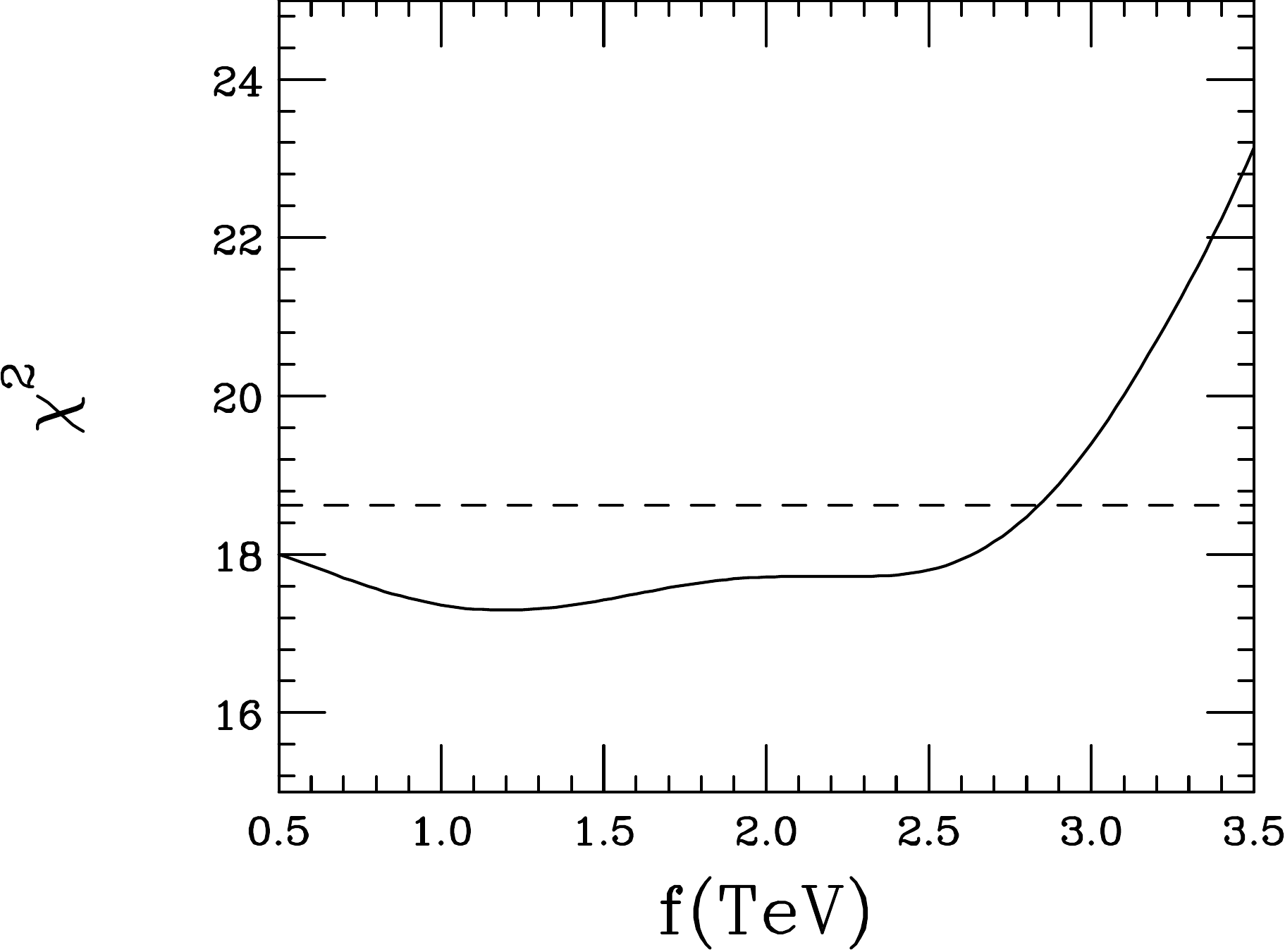}}
\caption{\chisqsp as a function of $f$ for the LH model. The 
  dashed line indicates the \chisqsp value of the SM best fit.}
\label{fig2}
\end{figure}

The 95\% CL contour in the $f - c^{\prime}$ plane is shown in figure
1. The dashed line is the trajectory of the best fit. As in
\cite{chkmt2,mss} the contour is defined with respect to the SM best
fit, although with a more restrictive criterion: we require $\Delta
\chi^2 < 5.99$ corresponding to 95\% CL for two degrees of freedom
($f$ and $c^{\prime\, 2}$), compared to $\Delta \chi^2 < 7.8$ and
$\Delta \chi^2 < 6.6$ in \cite{chkmt2} and \cite{mss} respectively.
The global best fit with $\chi^2 = 17.3$ is at $f=1.1$ TeV, 1.3
\chisqsp units below the SM best fit with $\chi^2 = 18.6$. As seen in
figure 2 the \chisqsp distribution as a function of $f$ is extremely
flat, varying by less than one \chisqsp unit for $f$ between 0.5 and
2.7 TeV.  The upper limit at $f=3.5$ TeV is a consequence of the fine
tuning constraint.  The fits prefer large values of the Higgs boson
mass, well above the 114 GeV LEPII lower limit. The \chisqsp 
distribution is also very flat as a function of \mh, as can be 
seen in figure 3 for $f=1.1$ TeV, where \chisqsp varies by no more 
than 0.2 units between $m_H=300$ GeV and $m_H=1$ TeV.

\begin{figure}[t]
\centerline{\includegraphics[width=5in,angle=0]{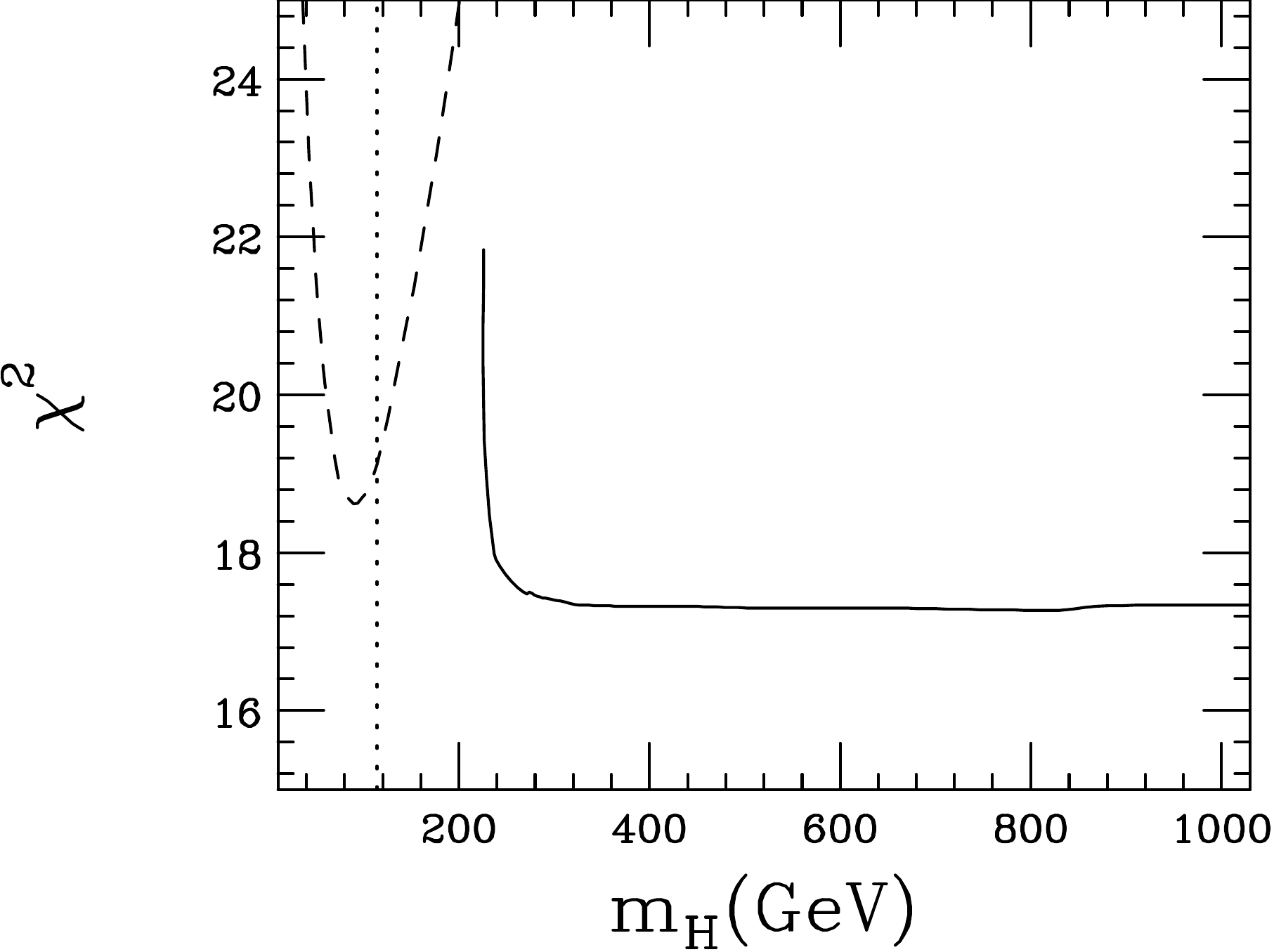}}

\caption{\chisqsp distributions as a function of \mhsp for the LH
  model with $f=1.1$ TeV (solid line) subject to the three boundary
  conditions and for the SM (dashed line). The dotted line is
  the LEPII lower bound on \mh.}

\label{fig3}
\end{figure}

These results are quite different from the earlier
studies. In \cite{chkmt2} fits with $f=1$ TeV are at the limit defined
by $\Delta \chi^2 = 7.8$, hence 9 \chisqsp units above the value
obtained here, nor do they satisfy the fine tuning constraint.  Those
fits only improve as $f$ increases, as the effects of the model
begin to decouple. In contrast the best fits presented here are at
$f\simeq 1$ TeV and the upper limit on $f$ is set by the fine tuning
constraint. The difference is principally the result of scanning on
the SM parameters, especially \mh, and on a more thorough scan over
the LH parameters including the $t-t^{\prime}$ mixing angle $s_L$. 

While the LH fit has more free parameters than the SM, the discovery
of a \zpsp boson in the EW allowed region would determine the
parameters $f$ and $c^{\prime}$, and the resulting LH fit would have a
comparable confidence level to the SM. Because of the nature of the SM
fit, the results obtained here are as good as it gets for any BSM
model that does not explicitly address the 3.2$\sigma$ \alrsp -
\afbbsp discrepancy with flavor-specific new physics, since all other
data agree as well or better than chance with the SM.\cite{mc-lose^2}
The large pull of \afbbsp is entirely responsible for the marginal
confidence level of the SM fit, as can be seen by comparing the SM
fits in tables 2 and 3.

The $b$ and $c$ quark asymmetry measurements, \afbbsp and \afbc, have
large QCD corrections that must be merged with the experimental cuts
using hadronic Monte Carlos, giving rise to a systematic uncertainty
that is difficult to estimate reliably.\cite{mc-lose^22} If they are
excluded the resulting SM fit is robust, with \chisqsp confidence
level increasing from $CL(18.6,13)=0.14$ for the full data set (table
2) to $CL(8.3,11)=0.69$ for the reduced set (table 3), but the central
value for the Higgs mass decreases from 89 GeV, with 24\% probability
to be in the LEPII allowed region above 114 GeV, to 58 GeV, with only
4\% probability for the allowed region.\cite{mc-lose^2,mc-lose^22} The
LH model raises the predicted value of the Higgs mass for the reduced
data set well above 114 GeV while maintaining the robust quality of
the fit to the EW data. The best fit occurs at $f=1.4$ TeV and
$m_H=520$ GeV with $\chi^2=8.0$. With discovery of a compatible \zpsp
this would imply a 53\% confidence level, $CL(8.0,9)=0.53$. For
$c^{\prime}= 0.38$ and $f=1.47$ TeV, corresponding to the CDF excess
at 240 GeV, the best fit has $\chi^2=8.46$ and a confidence level of
0.49. The \chisqsp of the best fit as a function of $f$ and \mhsp is
again quite flat as a function of \mh.  For both data sets the
\chisqsp minimum is nearly independent of \mhsp (for large enough \mh)
because shifts in \mhsp are compensated by shifts in the LH
parameters, especially $s_L$ and $v^{\prime}$.

\vskip 0.1 in
\noindent{\bf The Centi-Weak \zpsp Boson:}

A characteristic prediction emerging from the fits is a light, narrow
\zpysp boson, between $\sim 100$ and 500 GeV, coupled to SM
hypercharge $Y$. The EW fits favor values of $c^{\prime}$ near $\sqrt
\eta^{\prime}=\sqrt{0.4}$ which suppresses the strength of the
coupling and reduces the effect of \zzpsp mixing on EW 
observables, as can be seen in equation (1).  The mass is determined
by the LH model parameters $f$ and $c^{\prime}$,
$$
m_{Z^{\prime}}= \frac{s_W}{\sqrt{5}s^{\prime}c^{\prime}}\, 
                \frac{f}{v}\, m_Z,           \eqno{(6)}
$$ 
implying a light \zpsp boson, because of the factor $1/\sqrt{5}$ and
especially because the fits favor values of $c^{\prime}$ that maximize
the factor $s^{\prime}c^{\prime}$ in the denominator.  Neglecting
(for the moment) 
\zzpsp mixing, which is of order $v^2/f^2$, the \zp-fermion
interaction is
$$
{\cal
  L}_{Z^{\prime}ff}=g_{Z^{\prime}}\overline f y_f{ \slash\!\!\!\!
  {Z_Y}}^{\prime}f                                  \eqno{(7)}
$$        
with $y_f=q_f-t_{3f}$, where $g_{Z^{\prime}}$ is related to the SM $Z$
coupling $g_Z=g/{\rm cos}\theta_W$ by
$$
r_{Z^{\prime}}\equiv \frac{g_{Z^{\prime}}}{g_Z} =
  \frac{s_W(c^{\prime\,2}-\eta^{\prime})}{s^{\prime}c^{\prime}}.   \eqno{(8)}
$$ 
However, because $g_{Z^{\prime}}$ is suppressed for $c^{\prime\, 2}$
near $\eta^{\prime}=0.4$, the small admixture of the SM $Z_0$ boson in
the \zpsp mass eigenstate can have a significant effect on the
interaction of the \zp.  With $Z^{\prime}\simeq Z_Y -
\theta_{Z-Z^{\prime}}Z_0$ the \zzpsp mixing angle at leading order is
$$
\theta_{Z-Z^{\prime}} = 
       \frac{s_W(1-2c^{\prime\, 2})}{2s^{\prime}c^{\prime}}
       \frac{m_Z^2}{m_{Z^{\prime}}^2}.    \eqno{(9)}
$$ 
Including the effect of \zzpsp mixing on $Z^{\prime}ff$ interactions 
we replace equation (7) with
$$
{\cal
  L}_{Z^{\prime}ff}=g_Z\overline f g^{\prime}_f{ \slash\!\!\!\!
  {Z}}^{\prime}f                                  \eqno{(10)}
$$
where  
$$
g^{\prime}_f = r_{Z^{\prime}}y_f - \theta_{Z-Z^{\prime}}
                 (t_{3f} - s_W^2 q_f).         \eqno{(11)}
$$ 
In particular, the \zpsp eigenstate then has an appreciable
branching ratio to $W^+W^-$ when $r_{Z^{\prime}}$ is small.

Using equations (6) and (8-11) the \zpsp mass and couplings are
determined in terms of $f$ and $c^{\prime}$.  The cross section for
\zpsp production as a function of $m_{Z^{\prime}}$ is then determined
by the two parameters, $f$ and $c^{\prime}$. We compute the cross
sections\footnote{Collider cross sections are computed with Madgraph
  v4\cite{mgv4}.} with $K$ factor $K=1.3$ and compare the results
with the limits on narrow \zpsp production from CDF\cite{cdf} and
D0\cite{d0}.  An upper limit on $\sigma_{Z^{\prime}} BR(e^+e^-)$ at
mass $m_{Z^{\prime}}$ implies upper and lower limits on $c^{\prime\,
  2}$ at corresponding values of $f$, while a \zpsp discovery would
determine $f$ and $c^{\prime\, 2}$ up to a twofold ambiguity. Solving
numerically we obtain the upper and lower limits on $c^{\prime\, 2}$
as a function of $f$ that are shown in figure 1.

Viewing the CDF excess at 240 GeV as illustrative of a possible
signal,\footnote{Figure 1 of \cite{cdf} compares the data to a fit
  assuming a \zpsp at 241.3 GeV with negligible intrinsic width. The
  fit predicts events principally in two bins between 230 and 250 GeV,
  while most of the excess is in the upper bin. However, with only
  $\sim 30$ excess events no conclusion can be drawn from the shape of
  the distribution.}  we estimate $\sigma_{Z^{\prime}} BR(e^+e^-)
\simeq 42$ fb from the data.\footnote {The quoted net signal
  efficiency\cite{cdf} at $m_{e^+e^-}=150$ GeV is $\epsilon_{\rm TOT}
  = 0.27$. The net efficiency at 240 GeV increases in proportion to
  the acceptance of the CDF fiducial region, to $\epsilon_{\rm TOT} =
  0.32$, since trigger and other instrumental efficiencies for
  electrons in the fiducial region vary slowly between 150 and 240
  GeV. Taking the interval $m_{e^+e^-}=240\, {\rm GeV}\pm
  2\sigma_{m_{e^+e^-}}$ where the CDF resolution at 240 GeV is
  $\sigma_{m_{e^+e^-}}= 5.4$ GeV, we find 32 signal and 70 background
  events from figure 1 of \cite{cdf}, reproducing the quoted
  3.8$\sigma$ nominal signficance.  The total signal cross section for
  2.5 fb$^{-1}$ is then $\simeq 42$ fb.}  The corresponding values of
$f$ and $c^{\prime\,2}$ are plotted as circles in figure 1 at (1.47
TeV, 0.38) and (1.50 TeV, 0.43). The solution at $f=1.47$ TeV is
preferred by $\Delta \chi^2=4$, and the properties of the \zpsp boson
for this choice are displayed in table 1. Because
$r_{Z^{\prime}}=g_{Z^{\prime}}/g_Z=0.021$ and
$\theta_{Z-Z^{\prime}}=0.017$ are both small the \zpsp is extremely
narrow, with a width of 11 MeV.  Because $r_{Z^{\prime}}$ and
$\theta_{Z-Z^{\prime}}$ are comparable in magnitude the effect of
\zpsp mixing on the properties of the \zpsp is signficant. The
$e^+e^-$ branching ratio is 10\%, reduced by \zzpsp mixing from the
$\sim 15\%$ branching ratio of a \zpsp boson coupled to
hypercharge. There is a substantial 29\% branching ratio to $W^+W^-$
which is entirely due to the SM $Z_0$ component of the \zpsp mass
eigenstate. The $e^+e^-$ branching ratio and $W^+W^-$ decay can be
used to distinguish the LH \zpsp boson from other narrow \zpsp bosons
that couple predominantly to hypercharge, e.g., by kinetic
mixing\cite{holdom} or by the Stueckelberg mechanism.\cite{stklbg}

\begin{table}
\begin{center}
\vskip 12pt
\begin{tabular}{c|c}
\hline
\hline
$m_{Z^{\prime}}$& 240 GeV\\
$\sigma_{Z^{\prime}} BR(e^+e^-)$ & $\sim 42$ fb\\
\hline
$f$ & 1.47 TeV \\
$c^{\prime\, 2}$ & 0.38 \\
\hline
$r_{Z^{\prime}}=g_{Z^{\prime}}/g_Z$&0.021\\
$\theta_{Z-Z^{\prime}}$ & 0.017 \\
$\Gamma_{Z^{\prime}}$& 15 MeV\\
$BR(e^+e^-)$ & 0.10 \\
$BR(\overline \nu_e \nu_e)$ & 0.0004\\
$BR(\overline bb)$ & 0.067\\ 
$BR(\overline cc)$ & 0.10\\ 
$BR(W^+W^-)$ &  0.29 \\ 
\hline
\hline
\end{tabular}
\end{center}
\caption{Properties of the hypothetical \zpsp boson based on the CDF 
$e^+e^-$ excess at 240 GeV.}
\end{table}

The best fit with \zpsp parameters corresponding to the 240 GeV CDF
excess, as in table 1, has $\chi^2=17.4$, which is 1.2 units below the
SM best fit. Varying $m_H,s_L$, and $v^{\prime}$ there is a range of
fits with similar \chisqsp values, with \mhsp going from 270 GeV to 1
TeV and $\delta_{FT}$ from 0.1 to 1.3. One of these, with $m_H=820$
GeV and $\delta_{FT}=0.9$, is shown alongside the SM best fit in table
2. For this fit the masses of the top partner, triplet Higgs, and
heavy $W$ are 2.2, 7.2, and 2.1 TeV respectively. However these masses
are not well determined since other fits with comparable \chisqsp
values predict different masses.

\begin{table}
\begin{center}
\vskip 12pt
\begin{tabular}{c|c|cc|cc}
\hline
\hline
 &Experiment& {\bf SM} & Pull &{\bf LH} & Pull\\
\hline
$A_{LR}$ & 0.1513 (21)  & 0.1480  & 1.6& 0.1472 & 1.9 \\
$A_{FB}^l$ & 0.01714 (95) &0.01644  & 0.7&0.01626  &0.9 \\
$A_{e,\tau}$ & 0.1465 (33) & 0.1480 & -0.5& 0.1472 & -0.2 \\
$A_{FB}^b$ & 0.0992 (16) & 0.1038 & -2.9&0.1032&-2.5 \\
$A_{FB}^c$ & 0.0707 (35) & 0.0742 & -1.0&0.0737&-0.9 \\
$\Gamma_Z$ & 2495.2 (23) & 2495.7 &-0.2& 2496.6 & -0.6 \\
$R_{\ell}$ & 20.767 (25) &20.739  & 1.1 &20.741 & 1.0 \\
$\sigma_h$ & 41.540 (37) & 41.481 &1.6  & 41.478 &1.7 \\
$R_b$ & 0.21629 (66) & 0.21582 &0.7& 0.21561 &1.0 \\
$R_c$ & 0.1721 (30) & 0.1722 &-0.04& 0.1723 &-0.07 \\
$A_b$ & 0.923 (20) & 0.935 &-0.6 & 0.935 &-0.6 \\
$A_c$ & 0.670 (27) &  0.668 & 0.07&  0.668& 0.09 \\
$m_W$ & 80.399 (23) & 80.378 & 0.9 & 80.393 & 0.3 \\
$A_{PV}$ & $-131\, (17)\cdot 10^{-9}$ &$-156\cdot 10^{-9}$ &1.4&
           $-154\cdot 10^{-9}$&1.3  \\
$Q_W(Cs)$ & -73.16 (.35) &-73.14 &-0.06&-73.33&0.5  \\
\hline
$\Delta \alpha^{(5)}(m_Z)$ & 0.02758 (35) &0.02768& -0.3 & 0.2761 & -0.09\\
$m_t$ & 173.3 (1.1) &173.3  &0.02 &173.3  &0.02 \\
$\alpha_S(m_Z)$ &    &0.1180& &0.1198& \\
$m_H$ & & 89 && 820 \\
\hline
$c$ & & & &0.24 & \\
$v^{\prime}$ (GeV) & & & &1.0 & \\
$s_L$ & & & & 0.11& \\
\hline
$\chi^2$/dof& & 18.6/13 && 17.7/11 \\
CL($\chi^2)$ & & 0.135 &&0.09 \\
\hline
$m_H[90\%]$ (GeV) &  &51 --- 152 & &270 --- 1000 & \\
CL($m_H\ > 114$ GeV)& &0.24 & & 1  & \\
\hline
\hline
\end{tabular}
\end{center}
\caption{The SM best fit and an LH model fit with $f,c^{\prime\, 2}= 1.47
\, {\rm TeV}, 0.38$ corresponding to the CDF excess at 240 GeV.}
\end{table}

Because the Higgs triplet and the top partner both effect the EW fit
predominantly via the oblique parameter $T$, the fit sees $v^{\prime}$
and $s_L$ as a single degree of freedom.\footnote{However we vary
  $v^{\prime}$ and $s_L$ separately to verify the boundary conditions,
  equations (4-5).} Since a \zpsp discovery would determine $f$ and
$c^{\prime}$, the resulting LH fit would effectively have two more
independent parameters than the SM fit: the $SU(2)$ mixing angle $c$
and the oblique parameter $T$ determined in a correlated way by
$v^{\prime}$ and $s_L$. The LH fit to the full data set in table 2
then has 11 degrees of freedom with confidence level
$CL(17.7,11)=0.09$, comparable to the SM fit with $CL(18.6,13)=0.135$.
For the reduced data set shown in table 3 with the hadronic
asymmetries \afbbsp and \afbcsp omitted, the best LH fits have a
robust confidence level, $CL(8.6,9)=0.48$, as does the SM fit with
$CL(8.3,9)=0.69$. However, unlike the SM fit which predicts a 58 GeV
Higgs boson with only 4\% probability to be in the LEP II allowed
region above $114$ GeV, the LH fit has a shallow \chisqsp minimum at
$m_H=560$ GeV and is consistent with values between 270 GeV and 1 TeV.

\begin{table}
\begin{center}
\vskip 12pt
\begin{tabular}{c|c|cc|cc}
\hline
\hline
 &Experiment& {\bf SM} & Pull &{\bf LH} & Pull\\
\hline
$A_{LR}$ & 0.1513 (21)  & 0.1498  & 0.7& 0.1492 & 1.0 \\
$A_{FB}^l$ & 0.01714 (95) &0.01683 & 0.3&0.01670  &0.5 \\
$A_{e,\tau}$ & 0.1465 (33) & 0.1498 & -1.0& 0.1492 & -0.8 \\
$\Gamma_Z$ & 2495.2 (23) & 2496.4 &-0.5& 2496.6 & -0.6 \\
$R_{\ell}$ & 20.767 (25) &20.743  & 1.0 &20.741 & 1.0 \\
$\sigma_h$ & 41.540 (37) & 41.480 &1.6  & 41.480 &1.6 \\
$R_b$ & 0.21629 (66) & 0.21581 &0.7& 0.21573 &0.9 \\
$R_c$ & 0.1721 (30) & 0.1723 &-0.06& 0.1723 &-0.06 \\
$A_b$ & 0.923 (20) & 0.935 &-0.6 & 0.935 &-0.6 \\
$A_c$ & 0.670 (27) &  0.669 & 0.04&  0.668& 0.06 \\
$m_W$ & 80.399 (23) & 80.401 & -0.07 & 80.400 & -0.02 \\
$A_{PV}$ & $-131\, (17)\cdot 10^{-9}$ &$-159\cdot 10^{-9}$ &1.6&
           $-157\cdot 10^{-9}$&1.5  \\
$Q_W(Cs)$ & -73.16 (.35) &-73.09 &-0.2&-73.23&0.2  \\
\hline
$\Delta \alpha^{(5)}(m_Z)$ & 0.02758 (35) &0.02761& -0.09 & 0.2754 & 0.12\\
$m_t$ & 173.3 (1.1) &173.3  &0.02 &173.3  &0.02 \\
$\alpha_S(m_Z)$ &    &0.1180& &0.1186& \\
$m_H$ & & 58 && 560 \\
\hline
$c$ & & & &0.14 & \\
$v^{\prime}$ (GeV) & & & &0.5 & \\
$s_L$ & & & & 0.077& \\
\hline
$\chi^2$/dof& & 8.3/11 && 8.6/9 \\
CL($\chi^2)$ & & 0.69 &&0.48 \\
\hline
$m_H[90\%]$ (GeV) &  &30 --- 111 & &270 --- 1000 & \\
CL($m_H\ > 114$ GeV)& &0.04 & & 1  & \\
\hline
\hline
\end{tabular}
\end{center}
\caption{The SM best fit and an LH model fit with \afbbsp and \afbcsp
  excluded, the LH fit at $f,c^{\prime\, 2}= 1.47 \, {\rm TeV}, 0.38$
  corresponding to the CDF excess at 240 GeV.}
\end{table}

\vskip 0.1in
\noindent{\bf Future Prospects}

Future Tevatron and LHC data will determine if the excess at 240 GeV
is a real signal or, if not, can tighten the limits on $c^{\prime\,
  2}$ to the point of implausibility unless motivated by UV
completion of the model. As seen in figure 1 the difference between
the upper and lower limits on $c^{\prime\, 2}$ from the current CDF
and D0 searches ranges from 0.02 to 0.09 for $m_{Z^{\prime}}$ from
150 to 500 GeV. The limit scales with the integrated luminosity like
$\simeq L^{-\frac{1}{4}}$.  With $L=10\ {\rm fb}^{-1}$ the CDF and D0
limits could tighten by 30\% and 15\% respectively.

To illustrate the sensitivity of the LHC for the LH \zpsp boson we
consider $m_{Z^{\prime}}= 240$ GeV and $c^{\prime\, 2}= 0.38$,
corresponding to the CDF excess.  We require $p_T > 25$ GeV and
$|\eta| < 2.4$ for $e^+$ and $e^-$, and assume 65\% efficiency within
the fiducial region, as aready achieved by ATLAS in an early study of
the $Z$ boson\cite{atlasZ}. We parameterize the \epemsp fractional
mass resolution, $\hat\sigma_m = \sigma_m/m$, by $d_m=
\hat\sigma_m/0.02$, since $\hat\sigma_m \simeq 2\%$ for CDF at 240
GeV, a figure that will eventually be surpassed by ATLAS and CMS. The
signal region is defined as $m_{Z^{\prime}} \pm 2\, \sigma_m$. At
$\sqrt{s}=7$ TeV the CDF $3.8\sigma$ excess would then have a
significance of $9\sigma \cdot \sqrt{(L/d_m)}$ with $L$ expressed in
${\rm fb}^{-1}$. The current $40\, {\rm pb}^{-1}$ data sample is
inconclusive since it corresponds to only $1.8\, \sigma$ for $d_m=1$,
while the $\gtap\, 1{\rm\, fb}^{-1}$ sample projected for the coming
year can decisively exclude or confirm the predicted signal. For the
LHC at 7 TeV with $L=d=1$ the expected 95\% CL limit on production of
the 240 GeV LH \zpsp would imply $0.39 < c^{\prime\, 2} < 0.42$. At 14
TeV with $L=100$ and $d=1/2$ the expected 95\% constraint on \zpsp
would imply $0.40 < c^{\prime\, 2} < 0.41$. For the heaviest \zpsp
allowed by the EW fit, $m_{Z^{\prime}}=500$ GeV, the corresponding
95\% limits at 7 and 14 TeV would be $0.37 < c^{\prime\, 2} < 0.43$
and $0.39 < c^{\prime\, 2} < 0.41$.

The Tevatron \zpsp limits constrain the model for $m_{Z^{\prime}} \geq
150$ GeV, corresponding to $f\, \gtap\, 920$ GeV, but the region
within the 95\% EW contour with $f < 900$ GeV and $m_{Z^{\prime}}\,
<\, 150$ GeV is largely unconstrained. Good EW fits exist down to
$f=500$ GeV, corresponding to $m_{Z^{\prime}}=85$ GeV. The constraints
are relatively weak because LEPII ran sparsely below 150 GeV,
accumulating only $3\, {\rm pb}^{-1}$ samples at 130 and 136 GeV. For
instance, a 140 GeV \zpsp with $r_{Z^{\prime}}= 0.02$ corresponding to
$f\simeq 860$ GeV and $c^{\prime\, 2}=0.38$ (marked by the ellipse in
figure 1) yields a good EW fit to the full data set with $\chi^2 =
17.5$, $m_H= 600$ GeV, and $\delta_{FT}=1.1$. The resulting shift in 
$\sigma(e^+e^- \to \mu^+\mu^-)$ at 136 GeV is 0.22 fb, well below the
0.67 fb experimental uncertainty.\cite{lepii} Even at the 95\% limit
of the EW fit, $c^{\prime\, 2}=0.365$, the effect is only as big as
the experimental uncertainty.

For $f$ approaching 500 GeV the expansion in $v^2/f^2$ becomes
unreliable. Comparing the leading order result for
$\theta_{Z-Z^{\prime}}$ with the result to all orders\cite{hmnp} we
find that the corrections are $\leq O(10\%)$ for $f \geq 1$ TeV as
naively expected. \zzpsp mixing is kept under control despite the
light \zpsp mass, because the factor 1/5 that suppresses
$m_{Z^{\prime}}^2$ is cancelled by a factor $1-2c^{\prime\, 2}\simeq
1/5$ in the off-diagonal matrix element of the \zzpsp mass matrix. The
errors introduced by the leading approximation at the smallest values
of $f$ will shift the values of the parameters at which the best fits
occur but will not significantly alter the confidence levels. A
quantitatively reliable analysis of the very low $f$ region will
require going beyond the leading approximation.

\noindent{\bf Fine-tuning of model parameters}

While we have obtained fits to the EW data that resolve the little
hierarchy fine-tuning problem, we also find that recent Tevatron data
imposes a strong constraint on $c^{\prime}$, the $U(1)$ mixing angle
parameter. As shown in figure 1, the Tevatron data requires
$c^{\prime\, 2}$ to be near $\eta^{\prime}=2/5$ to suppress the \zpsp
coupling to SM fermions (see equation (8)). It is interesting that the
EW data requires no further fine-tuning ``price'': although simple
estimates suggest otherwise, we find that once the current Tevatron
constraint on $c^{\prime}$ is satisfied, no further tuning is
required.

To illustrate the extent of tuning required by the EW data we consider
the shift in $Z$-fermion couplings from \zzpsp mixing, $\delta g_f$,
and the corrections to the effective leptonic weak interaction mixing
angle, $x_W^{\ell,\, eff}$, which is the most important
pseudo-observable in the EW fit, with part per mil precision, $\delta
x_W^{\ell,\, eff}/x_W^{\ell,\, eff} \sim 1 \cdot 10^{-3}$. The
corrections to $\delta g_f$ from the heavy gauge bosons, $W^{\prime}$
and \zp, are shown in equation (1), while for $x_W^{\ell,\, eff}$ they
are
$$
\delta x_W^{\ell,\, eff}|_{W^{\prime},Z^{\prime}} = 
\frac{x_W(1-x_W)}{1-2x_W}\frac{v^2}{f^2}\left(-\frac{5}{4}
     +c^2(1-c^2) +5c^{\prime\, 2}(1-c^{\prime\, 2})\right). \eqno{(12)}
$$ 
The factor 5 amplifying the $U(1)$ corrections in equations (1) and
(12) is especially dangerous. 

As a specific example we consider the fit in table 2 of the CDF excess
at 240 GeV. The prefactors $v^2/2f^2$ and $\simeq v^2/3f^2$ in
equations (1) and (12) are then of order $0.01$, while the factors in
parentheses containing the $c$ and $c^{\prime}$ dependence are
generically of order one, suggesting that fine tuning is required.
This is indeed the case but the necessary tuning is already imposed by
the Tevatron \zpsp bounds. Notice first that in equation (1) the
$U(1)$ correction is suppressed not only by the factor $(c^{\prime\,
  2} - 2/5)$ but, as an added bonus, the dangerous factor 5 is offset by
the factor $(1 - 2c^{\prime\, 2})$ which is $\simeq 1/5$ if
$c^{\prime\, 2}\simeq 2/5$.  The $SU(2)$ correction is also small, as
the EW fit prefers small values for $c$, typically between 0.1 and
0.3, with for instance $c=0.24$ for the fit in table 2. The net result
for the correction to the $Zee$ coupling is then $\delta g_e \simeq
2\cdot 10^{-4}$, two orders of magnitude below the naive estimate 
and within the range of the experimental uncertainty. A
similar miracle occurs for $\delta x_W^{\ell,\, eff}$, equation (12),
where the term $-5/4$ is offset by the $c^{\prime}$ dependent term,
(e.g., at $c^{\prime\, 2} = 2/5$ we have $-5/4 + c^{\prime\, 2}(1 -
c^{\prime\, 2}) = -1/20$) with some further reduction provided by the
$SU(2)$ term. The net result, $\delta x_W^{\ell,\, eff} \simeq 1.7
\cdot 10^{-4}$ is again reduced by two orders of magnitude from the
naive estimate and falls within the precision of the measurement.  The
suppression of these (and the other) EW corrections is assured just by
the value of $c^{\prime}$ imposed by the Tevatron data with no
additional fine tuning.

Corrections from the $T$ parameter due to the top partner,
$t^{\prime}$, and the triplet Higgs vacuum expectation value,
$v^{\prime}$, can also have an appreciable effect on the fit, but they
are not fine-tuned and they are not solely or even primarily
responsible for the preference for large values of \mhsp in the
fits. For instance, for $f$ and $c^{\prime}$ fixed by CDF data as in
table 2, there is an acceptable fit with $T_{t^{\prime}} =
T_{v^{\prime}}= 0$, falling within the 95\% CL contour of figure 1,
with $\chi^2=20.1$, $\delta_{FT}= 0.11$, and $m_H=700$ GeV.  Allowing
$T_{t^{\prime}} = 0.11$ and $T_{v^{\prime}}= 0.01$, the fit in table 2
improves to $\chi^2=17.7$, $\delta_{FT}= 0.9$ at $m_H=820$ GeV. These
values are not fine-tuned, as a broad range of other values of
$T_{t^{\prime}}$ and $T_{v^{\prime}}$ also provide robust (and in some
cases slightly better) fits, with a wide range of \mhsp values,
between 300 GeV and 1 TeV. The values of $T_{t^{\prime}}$,
$T_{v^{\prime}}$, and \mhsp are indeed correlated but they are not
fine tuned. Different values of $T_{t^{\prime}}$ and $T_{v^{\prime}}$
can be accomodated with different choices of \mh. The correlations 
imply predictions that will be tested if evidence for the model 
emerges and the model parameters are measured. 

\noindent{\bf Discussion}

Contrary to earlier studies we find that the original littlest Higgs
model with universal fermion couplings is consistent with precision EW
data while amelioraing the little hierarchy problem, as originally
envisioned in \cite{ackn}.  Our conclusions differ from the earlier
studies\cite{chkmt2,mss} chiefly because we have scanned over all SM
and LH parameters, as might not have even been possible for the
earlier studies with the computing capability available at the time.
However, in the intervening years the Tevatron limits on \zpsp
production have increased to the point that they now constrain the
model more strongly than the EW data. The excess observed by CDF at
240 GeV is consistent with the predicted \zpsp boson; if it is
confirmed the LH model will be one of the possible explanations. If it
is not confirmed and no other consistent signal is seen at the
Tevatron or the LHC, the model will succumb to a fine tuning problem
for the $U(1)$ mixing angle $c^{\prime}$ that is as severe as the
tuning required by the little hierarchy problem that it purports to
solve.

\vskip .1in
\noindent {\it Acknowledgements:} 
{\small This work was supported in part by the Director, Office of
Science, Office of High Energy and Nuclear Physics, Division of High
Energy Physics, of the U.S. Department of Energy under Contract
DE-AC02-05CH11231}


\end{document}